\documentclass[preprint,aip,superscriptaddress,preprintnumbers,amsmath,amssymb]{revtex4-1}
\usepackage{amssymb}
\usepackage{amsmath}
\usepackage{graphicx}
\usepackage{multirow}
\usepackage{comment}
\usepackage{longtable}

\newcommand{\eqn}[1]{\mbox{Eq.\hspace{1pt}(\ref{#1})}}
\newcommand{\eqs}[2]{\mbox{Eq.\hspace{1pt}(\ref{#1}--\ref{#2})}}

\newcommand{\eqtn}[2]{\begin{equation} \label{#1} #2 \end{equation}}
\newcommand{\func}[1]{#1 \left[ \rho \right] }
\newcommand{\mfunc}[2]{#1_{#2} \left[ \rho \right] }

\newcommand{\pot}[1]{v_{\rm #1}}

\def\brp{{\mathbf{r}^{\prime}}}

\def\br{{\mathbf{r}}}

\def\d{{\mathrm{d}}}

\def\se{{Schr\"{o}dinger equation}}

\begin{document}
\title{Nonlocal Kinetic Energy Functionals By Functional Integration}

\author{Wenhui Mi}
\affiliation{Department of Chemistry, Rutgers University, Newark, NJ 07102, USA}
\author{Alessandro Genova}
\affiliation{Department of Chemistry, Rutgers University, Newark, NJ 07102, USA}
\author{Michele Pavanello}
\email{m.pavanello@rutgers.edu}
\affiliation{Department of Chemistry, Rutgers University, Newark, NJ 07102, USA}

\begin{abstract}
{Since the seminal works of Thomas and Fermi, researchers in the Density-Functional Theory (DFT) community are searching for accurate electron density functionals. Arguably, the toughest functional to approximate is the noninteracting  Kinetic Energy, $T_s[\rho]$, the subject of this work. The typical paradigm is to first approximate the energy functional, and then take its functional derivative, $\frac{\delta T_s[\rho]}{\delta \rho(\br)}$, yielding a potential that can be used in orbital-free DFT, or subsystem DFT simulations. Here, this paradigm is challenged by constructing the potential from the second-functional derivative {\it via} functional integration. A new nonlocal functional for $T_s[\rho]$ is prescribed (which we dub MGP) having a density independent kernel. MGP is constructed to satisfy three exact conditions: (1) a nonzero ``Kinetic electron'' arising from a nonzero exchange hole; (2) the second functional derivative must reduce to the inverse Lindhard function in the limit of homogenous densities; (3) the potential derives from functional integration of the second functional derivative. Pilot calculations show that MGP is capable of reproducing accurate equilibrium volumes, bulk moduli, total energy, and electron densities for metallic (BCC, FCC) and semiconducting (CD) phases of Silicon as well as of III-V semiconductors. MGP functional is found to be numerically stable typically reaching selfconsistency within 12 iteration of a truncated Newton minimization algorithm. MGP's computational cost and memory requirements are low and comparable to the Wang-Teter (WT) nonlocal functional or any GGA functional.}
\end{abstract}
\maketitle
\newpage

\section{Introduction}
\subsection{Orbital-free and Kohn--Sham DFT}
Since the seminal work of Kohn and Sham \cite{kohn1965}, the energy functional of the electron density can be written in terms of the noninteracting Kinetic energy (KE) functional, $T_{\rm s}[\rho]$, and the associated exchange-correlation (xc) functional, $E_{\rm xc}[\rho]$, as follows
\eqtn{eq:ef}{E[\rho] = T_{\rm s}[\rho] + E_{\rm H}[\rho] + E_{\rm xc}[\rho] + E_{\rm eN}[\rho] +E_{\rm NN},}
The nuclear-nuclear interaction, $E_{\rm NN}$ is density independent. The electron-nuclear interaction $E_{\rm eN}[\rho]=\int v_\text{ext}(\br)\rho(\br)\d\br$, is linear in the density, and the other terms [for closed-shell systems and introducing the Kohn--Sham orbitals, $\{\phi_i[\rho](\br)\}$, including the yet unknown exchange--correlation (xc) energy density, $\varepsilon_{\rm xc}[\rho](\br)$], take the form
\begin{align}
\label{ts}
T_{\rm s}[\rho]   &= -\frac{1}{2}\sum_{i=1}^{N_e/2} \langle \phi_i[\rho] | \nabla^2 | \phi_i[\rho] \rangle, \\
\label{eh}
E_{\rm H}[\rho]  &=\frac{1}{2}\int\int \frac{\rho(\br)\rho(\brp)}{|\br - \brp|}\d\br\d\brp, \\ 
\label{ex}
E_{\rm xc}[\rho]&=\int\varepsilon_{\rm xc}[\rho](\br)\rho(\br)\d\br.
\end{align}

There are two different and theoretically equivalent ways to find the electron density associated to a given external potential. One prescribes the direct minimization of the DFT Lagrangian,
\eqtn{eq:lag}{\mathcal{L}[\rho]=E[\rho]-\mu\left( \int\rho(\br)\d\br - N_e \right),}
which is known as Orbital-Free DFT (OF-DFT). The other way (due to Kohn and Sham \cite{kohn1965}) prescribes solving the following \se
\eqtn{eq:KS}{-\frac{1}{2}\nabla^2\phi_i[\rho](\br) + \left[ \underbrace{\pot{H}[\rho](\br)+\pot{xc}[\rho](\br)+\pot{ext}(\br)}_{\pot{s}(\br)} \right] \phi_i[\rho](\br) = \varepsilon_i \phi_i[\rho](\br).}

\subsection{OF-DFT: The Kinetic energy functional conundrum}
If one decides to carry out the density search according to the minimization of the DFT Lagrangian in \eqn{eq:lag}, then there is no need to invoke the concept of Kohn--Sham (KS) orbitals. Instead, the electron density, $\rho(\br)$, can be utilized as the only variational parameter. The drawback is that in addition to approximations needed for the xc functional, in OF-DFT approximations for the Kinetic energy density functional (KEDF), $T_{\rm s}[\rho]$, are also needed.

Local and semilocal (i.e., dependent on the value of the electron density and its gradient at a point in space) parametrizations of KEDFs have shown potential \cite{Karasiev_2015,Trickey_2009,Karasiev_2009} and display a more than satisfactory agreement with KS-DFT for simulations of warm dense matter \cite{Karasiev_2013}. Recent work by the Della Sala group \cite{lari2014} advocates for employing the Laplacian in addition to the gradient in the formulation of $T_{\rm s}[\rho]$ , and the possibility of reproducing accurate Kinetic energy densities \cite{Smiga_2017}.  

Nonlocal versions of $T_{\rm s}[\rho]$, such as the Wang-Govind-Carter (WGC) \cite{wang1998}, the one proposed by Perrot \cite{perr1994}, the latest Huang-Carter (HC)\cite{huan2010}, and others \cite{chai2007}; typically improve over local and semilocal functionals \cite{xia2012,xia2012b,huan2010,ho2008,zhou2005,carl2003} and all share the following form
\begin{equation}
\label{tkin}
\func{T}=\mfunc{T}{\rm TF}+\mfunc{T}{\rm vW}+\mfunc{T}{\rm NL},
\end{equation}
where $\mfunc{T}{\rm TF}$ is the Thomas-Fermi Kinetic Energy \cite{fermi1927,thom1927}, $\mfunc{T}{\rm vW}$ is the gradient correction due to von Weizs\"acker \cite{weiz1935}, and $\mfunc{T}{\rm NL}$ is the remaining contribution which is by construction of nonlocal character.
The general form of $T_{NL}$ is
\begin{equation}
\label{tnadd1}
\mfunc{T}{\rm NL}=\int  \rho^\alpha\left( \br \right) 
\omega_{T_{NL}}\left[ \rho \right]\left( \br, \brp \right) \rho^\beta \left( \brp \right)\d\br\d\brp,
\end{equation}
where the function $\omega_{T_{NL}}$ is commonly called ``kernel of the nonlocal KEDF'', and $\alpha$ and $\beta$ are suitable exponents. 

Unfortunately, the current state of the art for OF-DFT is that, although the algorithms have dramatically improved over the years and are typically orders of magnitude faster than KS-DFT, their application to semiconductors and molecular systems is still problematic \cite{zhou2005,huan2010,xia2012,xia2012b}. For example,  WGC \cite{wang1998,wang1999,zhou2005} and WT \cite{wang1992} KEDFs can reproduce remarkably accurate bulk properties for light metallic systems, but are less accurate for semiconductors and nonsimple metals. The HC KEDF \cite{huan2010} produces generally reliable bulk properties for semiconductors. However, it underestimates the electron density in bonding regions of CD Silicon and III-V semiconductors, it is less accurate than WT for metals, and it is substantially more computationally expensive than any other KEDFs.\cite{huan2010,shin2014} Density decomposition schemes that combine nonlocal with other functionals have found applicability to molecules, materials, and alloys \cite{xia2012,shin2014,Xia_2016} and thus appear to be an interesting avenue of research. 

In this work, we aim at formulating a KEDF that is: (1) at least as accurate as WGC and WT KEDFs for nonsimple metals such as the BCC and FCC phases of Silicon; (2) at least as accurate as the HC KEDF in predicting bulk properties of CD Silicon and III-V semiconductors, especially in regards to the selfconsistent electron density in bonding regions; (3) computationally inexpensive. 

\section{Theoretical background}
It is known that the exact $T_s$ functional satisfies a number of mathematical relations also known as exact conditions. It is arguable \cite{perd2005} that the more exact conditions a functional satisfies, the closer that functional necessarily is to the exact functional. It is through imposition of exact conditions that the Trickey group developed GGA KE functionals approaching the accuracy of KS-DFT in the limit of high temperature \cite{Karasiev_2013}. By imposing exact conditions on the asymptotic behavior of the linear response function, the Carter group formulated the HC KEDF which could be applied to systems with record inhomogenous density for OF-DFT simulations \cite{huan2010}.  

Let us enumerate three exact conditions we believe are particularly important:  (1) existence of the ``Kinetic electron'', (2) hypercorrelation, i.e., potentials are related by functional integration to their respective second functional derivative; (3) Lindhard response in the Free Electron Gas (FEG) limit.
\subsection{The Kinetic electron}
The first unconstrained functional derivative of the DFT Lagrangian with respect to the density yields the Euler equation of DFT \cite{Liu_2004}. Namely,
\eqtn{eq:euler}{\pot{T_s}[\rho](\br)+\pot{H}[\rho](\br)+\pot{xc}[\rho](\br)+\pot{ext}(\br)=\mu.}
By taking the negative of the Laplacian of each term and rearranging, $-\nabla^2 \pot{T_s}[\rho](\br)=\nabla^2\pot{H}[\rho](\br)+\nabla^2\pot{xc}[\rho](\br)+\nabla^2\pot{ext}(\br)$.
After solving the associated Poisson equations for each term in the long-range limit \cite{leeu1994}, we find,
\eqtn{eq:lump2}{4\pi\rho_{\rm T_{\rm s}} (\br) = -4\pi\rho (\br) -4\pi\rho_{\rm xc} (\br)-4\pi\rho_N (\br),}
where $\rho_N$ is the charge density of the nuclei (ions) and $\rho_{\rm xc}$ is related to the exchange-correlation hole. At long ranges, $\rho_{\rm xc}$ is identical to the xc hole \cite{leeu1994}. Thus, at long ranges, $\rho_{\rm T_{\rm s}}$ is the Kinetic energy equivalent of the xc hole. However, upon simplification of the $4\pi$ terms and integration over all space, we find that 
\eqtn{eq:lump3}{\int \rho_{\rm T_{\rm s}} (\br) \d\br = - N_e + 1 + N_N = 1 + \text{total charge},} 
Where the integration of the negative of the xc hole is 1, of the electron density is $N_e$, and of the nuclear density is $N_N$, and the total charge of the system is defined as $N_N-N_e$.

We define $\rho_{\rm T_{\rm s}}$ as \textbf{the Kinetic electron}, as for neutral systems it integrates to $+1$, in contrast to the xc hole which integrates to $-1$.

\subsection{Hypercorrelation {\it via} a line integral}
It is possible to recover an entire functional from the corresponding functional derivative using the following line integral \cite{vanle1995,gaid2009}:
\begin{equation}
\label{eq:sta1}
E[\rho] = \int_0^1 dt \int v[\rho_t](\br) \frac{\d\rho_t(\br)}{\d t} \d\br,
\end{equation}
where we have chosen the linear interpolation path which we describe below. 

As the potential is the functional derivative $v(\br) = \frac{\delta E[\rho]}{\delta \rho(\br)}$, the result of the integration is path-independent. The scaled density, $\rho_t$, can then have almost any form \cite{vanle1995}, the simplest of which is given by a linear interpolation between the vacuum and the density,
\begin{equation}
\label{eq:sta2}
\rho_t(\br) = t \rho(\br)  ~~\rightarrow~~ \frac{\d \rho_t}{\d t} = \rho(\br).
\end{equation}

Thus, the $t$-integral equation reduces to,
\begin{equation}
\label{eq:sta3}
E[\rho] = \int \d\br \int_0^1 dt ~ v[\rho_t](\br) \rho(\br) . 
\end{equation}

The previous integral defines the energy density in terms of the hypercorrelated potential 
\eqtn{eq:hyp1}{\epsilon[\rho](r)=\int_0^1 dt ~ v[\rho_t](\br).}
Hypercorrelation is a term coined by Burke \cite{burk1998} to emphasize the fact that the energy density is directly related to the physical (correlated) potential evaluated for an array of electron densities ranging from the vacuum to the true density. 

The above equation can be considered yet another exact constraint any functional (including the Kinetic energy) should satisfy. 

Hypercorrelation does not only concern the energy density, but also the potential. That is, the potential must be related to the second functional derivative by
\eqtn{eq:hyp2}{v[\rho] (\br) = \int \d\brp \int_0^1 \d t  \frac{\delta v[\rho](\br)}{\delta\rho_t(\brp)} \rho(\brp) = \int \d\brp \int_0^1 \d t \frac{\delta^2E}{\delta\rho(\br)\delta\rho_t(\brp)}\rho(\brp).}
It is important to notice that in the definition of the scaled second functional derivative on the rhs,  $\rho_t$ appears only on one of the two functional derivatives.


In light of \eqn{eq:hyp2}, \eqn{eq:sta3} can be recast in terms of the second functional derivative as
\eqtn{eq:energy_final}{E[\rho] = \int \d\br \int \d\brp \int_0^1 \d t \int_0^1  \d t^{\prime} \frac{\delta^2E}{\delta\rho_{t}(\br)\delta\rho_{t^{\prime}}(\brp)}\rho(\br)\rho(\brp).}
We should remark that due to the derivative discontinuity \cite{Perdew1982}, the line integration in the above equations will cross discontinuities in the integrands when integrating the kernel to yield a potential, and also when integrating the potential to obtain an energy value. These discontinuities are integrable and do not pose formal problems.

\subsection{FEG response}
An exact condition for a KE functional is its relationship to the linear response function of the FEG. There is an established relation between the KE functional and the linear response function \cite{wang2000,Pearson_1993,hohe1964}. Taking a functional derivative of each term in the Euler equation \eqn{eq:euler}, 
\eqtn{eq:FEG}{\frac{\delta^2 T_{\rm s}}{\delta \rho(\br) \delta \rho(\brp)}=-\frac{\delta v_s(\br)}{\delta \rho(\brp)},}
and the response function is given by $\chi_s(\br,\brp)=\bigg[\frac{\delta v_s(\br)}{\delta \rho(\brp)}\bigg]^{-1}$. Unfortunately, the above relationship is only valid at self-consistency and trying to impose it directly would lead to impractical algorithms. However, for the FEG we have a simplified relationship, $\chi_s(\br,\brp) = \chi_s(|\br-\brp|)$, resulting in a one-dimensional Fourier transform, $\chi_\text{Lind}(\eta)$, the Lindhard function \cite{jone1971,Lindhard_1954},
\begin{equation}
\label{eq:sta11}
\chi_\mathrm{Lind}(\eta) = - \frac{k_F}{\pi^2} \left( \frac{1}{2} + \frac{1-\eta^2}{4\eta} \ln \left| \frac{1+\eta}{1-\eta} \right| \right),
\end{equation}
with $\eta=\frac{q}{2k_F}$ and the Fermi wavevector is given by, $k_F=(3\pi^2\rho_0)^{1/3}$, with $\rho_0=\frac{N_e}{V}$.

Thus, in the limit of a constant and periodic electron density, we must impose
\eqtn{eq:FEG2}{\hat{F}\bigg[\frac{\delta^2 T_{\rm s}}{\delta \rho(\br) \delta \rho(\brp)}\bigg]_{\rho(\br)=\rho_0}=-\frac{1}{\chi_\text{Lind}(\eta)}=\frac{\pi^2}{k_F}G_\mathrm{Lind}(\eta),}
where $\hat{F}\left[\cdot\right]$ stands for Fourier transform in $q$, the conjugate variable of $|\br-\brp|$. This exact condition is at the core of the original Hohenberg and Kohn functional \cite{hohe1964} and inspired the entire class of nonlocal KE functionals, nicely summarized in Ref.~\citenum{wang2000}. The function $G_\mathrm{Lind}(\eta)$ is introduced here,
\eqtn{eq:glind}{G_\mathrm{Lind}(\eta)=\left( \frac{1}{2} + \frac{1-\eta^2}{4\eta} \ln \left| \frac{1+\eta}{1-\eta} \right| \right)^{-1}.}
\section{Kinetic energy functional by integration}
\subsection{Imposing the response of the Free Electron Gas through hypercorrelation}
We cast our developments in terms of \eqn{tnadd1}. Particularly, in the following we will only focus on the nonlocal term of the KEDF.
We begin by espousing the idea that in the limit of modeling the Free Electron Gas (FEG) a nonlocal KEDF should recover the exact FEG linear response function. This can be imposed by hypercorrelation, \eqn{eq:hyp2}. Namely, 
\begin{equation}
\label{eq:sta6}
\frac{\delta T_{NL}[\rho]}{\delta\rho(\br)}=\pot{T_ {NL}}[\rho](\br) =  \int \d\brp \int_0^1 \d t \, \frac{\delta^{2}T_{NL}}{\delta\rho(\br)\delta\rho_t(\brp)} \rho(\brp),
\end{equation}
where $\frac{\delta\pot{T_{NL}}[\rho_{t}](\br)}{\delta \rho_{t}(\brp)} = \frac{\delta^2 T_{NL}[\rho_{t}]}{\delta \rho(\br) \delta \rho_{t}(\brp)}$ will be dependent on $\rho_t$ and consequently on $t$. Applying a local density approximation (LDA) on the polynomial terms of \eqn{eq:FEG2}, $\frac{\delta^2 T_{NL}}{\delta \rho(\br) \delta \rho_t(\brp)}$ can be rewritten as:
\begin{equation}
\label{eq:vt}
\frac{\delta^{2}T_{NL}}{\delta\rho(\br)\delta\rho_t(\brp)}\simeq \frac{\pi^2}{(3\pi^2)^{\frac{1}{3}}}\rho^{-\frac{1}{6}}(\br)G_{NL}[t\rho_{0}](|\br-\brp|)\left(t\rho\right)^{-\frac{1}{6}}(\brp),
\end{equation}
where, following the prescription of \eqn{tkin}, and \eqs{eq:sta11}{eq:FEG2}, the Fourier transform of $G_{NL}(|\br-\brp|)$ is taken to be
\begin{align}
G_{NL}(\eta)=G_\mathrm{Lind}(\eta) -3\eta^{2}-1.
\end{align}

In the above, we have substituted $\rho_0 \rightarrow t\rho_0$, which gives the following Fermi wavevector $k_F(t) = t^{\frac{1}{3}} \left(3\pi^2 \rho_0\right)^{\frac{1}{3}}$, and the reciprocal space variable is $\eta({q},t) = \frac{q}{2k_F(t)} = \frac{q}{2t^{\frac{1}{3}} \left(3\pi^2 \rho_0\right)^{\frac{1}{3}}}$.
Thus, substituting \eqn{eq:vt} into the \eqn{eq:sta6}, the nonlocal Kinetic potential,$v_{T_{NL}}[\rho](\br)$, can be rewritten as:
\begin{align}
v_{T_{NL}}[\rho](\br)= \frac{\pi^{2}}{(3\pi^{2})^{1/3}}\rho^{-\frac{1}{6}}(\br)\int d\brp \rho^{\frac{5}{6}}(\brp)\int_{0}^{1} \d t~ t^{-\frac{1}{6}}G[t\rho_{0}](\br,\brp). 
\end{align}
Once again, we point out that the integral above should cross discontinuities in the integrand as the density scaling parameter, $t$, is varied. This does not directly affect the integral above because the Lindhard function is continuous across all values of $t$.

Finally, the nonlocal Kinetic potential can be obtained by 
\begin{align}
\label{eq:fvnl}
\pot{T_{NL}}(\br) = \rho^{-1/6}(\br)\hat{F}^{-1} \left[\hat{F}\left[ \rho^{5/6}(\br) \right](q) \cdot \omega_{T}(q)\right].
\end{align}
Where the kernel in reciprocal $\omega_{T}(q)$ is obtained from integration by parts
\begin{align}
\label{eq:omega0}
\omega_{T}(q)&=\frac{6}{5}\frac{\pi^{2}}{(3\pi^{2})^{1/3}}\left(G_{NL}(\eta(q))-\int_{0}^{1}dt t^{\frac{5}{6}}\frac{d G_{NL}(\eta(q,t))}{dt}\right)\nonumber \\
&=\omega_{WT}(q)-\frac{6}{5}\frac{\pi^{2}}{(3\pi^{2})^{1/3}}\int_{0}^{1}dt ~t^{\frac{5}{6}}\frac{d G_{NL}(\eta(q,t))}{dt}.
\end{align}

The new kernel, $\omega_{T_{NL}}$, is given by the kernel of the WT KEDF \cite{wang1992} plus a correction term. We should point out that this correction term derives from assuming that the inverse Lindhard function is dependent on the electron density (this allowed us include in the line integral the average density $\rho_0$). Thus, we expect $\omega_{T_{NL}}$ in \eqn{eq:omega0} to be most appropriate for substantially nonuniform densities. For light (simple) metallic systems, the distribution of electron density is very close to the uniform electron gas, thus WT kernel is already very good for these systems. We will see in our pilot calculations that for more complex systems, such as (nonsimple) metallic phases of Silicon, IV and III-V semiconductors, the correction term in conjuction with an additional correction arising from the presence of the ``Kinetic electron'' remarkably improves upon the performance of WT.

We also note that in going from \eqn{eq:sta6} to \eqn{eq:vt} we have assumed that the inverse Lindhard function is a function of $t$ multiplied by $\rho_0$. It is arguable that symmetry properties of the second functional derivative should be imposed. In the supplementary document, we carry out a detailed analysis of symmetrization of the kernel. There, we propose an arithmetic symmetrization as well as a geometric symmetrization of the kernel. We note that in either cases, the resulting kernels yield results that are equivalent to the ones obtained with the kernel in \eqn{eq:omega0} ({\it vide infra}). 

\subsection{Imposing a nonzero Kinetic electron}
\eqn{eq:lump2} implies that the KE potential in reciprocal space is given by
\eqtn{eq:el1}{\tilde v_{T_s}(q) = \frac{4\pi}{q^2}\tilde \rho_{T_s}(q) \rightarrow \tilde \rho_{T_s}(q) = \frac{1}{4\pi} q^2 \tilde v_{T_s}(q),}
where the tilde indicates quantities which have been Fourier transformed in reciprocal space (the Fourier transform is kept one dimensional for sake of simplicity). Unfortunately, the potential derived by  \eqn{eq:fvnl} is zero in the low $q$ limit (or at long ranges in real space). That is, 
\eqtn{eq:lim}{\lim_{q\to 0} \tilde v_{T_s}(q)=\text{Constant},}
and with that the Kinetic electron is absent. 

\begin{figure}[htp]
\begin{center}
\includegraphics[width=0.8\textwidth]{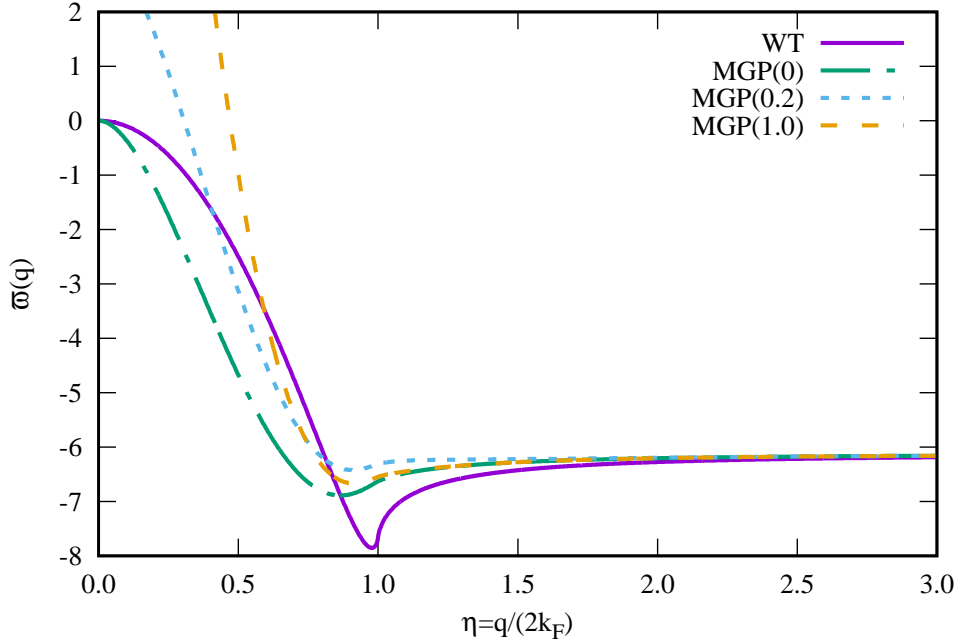}
\end{center}
\caption{\label{kernels} Comparison of four kernels: $\omega(q)$  on the $y$-axis, $q$ on the $x$-axis in atomic units. MGP(0) is the kernel from \eqn{eq:el2} without Kinetic electron. MGP(0.2) and MGP(1.0) are the kernels with the Kinetic electron parameter $a=b=0.2$ and $a=b=1.0$, respectively.  WT is the Wang--Teter kernel (Thomas--Fermi and von Weizs\"aker kernels removed from the inverse Lindhard function). All kernels are evaluated for $\rho_0=0.03~e^-/\text{Bohr}^3$ (yielding a value of $k_F\sim 1$) which is a typical value for valence electrons in bulk systems.}
\end{figure}

Thus, to impose a nonzero Kinetic electron, we decided to model it with a simple Gaussian multiplied by the square of the error function centered at $q=0$ in order to remove the Coulomb singularity. This results in a new kernel, $\omega_{MGP}$, obtained simply by modifying the expression for the kernel in \eqn{eq:omega0}. Namely,
\eqtn{eq:el2}{\omega_{MGP}(q)=\omega_{T_{NL}}(q)+{\rm erf}^{2} (q)\frac{4\pi a}{q^{2}}e^{-bq^{2}}.}
Although modeling the Kinetic electron by a simple Gaussian function is a gross approximation, in the future we commit to explore more accurate expressions inspired by established work of others \cite{perd1985,engel_dreizler}. We are aware that the exchange hole (and thus also the Kinetic electron) is generally not spherical \cite{beck2007}, thus in the future we will also consider nonspherical parametrizations.
In Figure \ref{kernels}, four different kernels are plotted. It is clear that the presence of the Kinetic electron imposed through \eqn{eq:el2} affects dramatically the low-$q$ limit of the kernel.  The kernel resulting from our manipulations is smooth, which will result in smooth nonlocal Kinetic energy potentials.

To understand the effect of the Kinetic electron, let us consider a very narrow Gaussian ($b\to +\infty$). This results in a spatially extended real-space Kinetic electron. This is the preferred shape for materials with a finite gap in which the dielectric screening is small. The opposite case ($b\to 0$) is preferred for metals, as there the dielectric screening is large and the real-space spatial extension of the Kinetic electron is reduced.  Thus, we expect to need smaller $b$ values to model metallic systems, and large $b$ values to model semiconductors and insulators. These observations are consistent with the finding \cite{Pick_1970,Adler_1962,Kim_2002,Nazarov_2011} (also exploited by Huang and Carter \cite{huan2010}) that, for $q\to 0$, the static response function for systems with gap, $\chi \to -\text{Constant}\times {q^2}$. An in-depth analysis of the relation between the low-$q$ behavior of $\chi$ and the long-range properties of the potential will be the subject of a follow-up work. 

\subsection{Introducing the MGP functional: details of the implementation }
We implemented the new functional resulting from the kernel in \eqn{eq:el2} according to the framework outlined in \eqn{eq:fvnl}.   
The MGP potential and energy functional are defined as
\begin{align}
\label{potential}
v_{T_s}(\br) =& v_{TF}(\br) + v_{vW}(\br) + \underbrace{\rho(\br)^{-\frac{1}{6}}\hat{F}^{-1}\left[ \hat{F}\left[ \rho^{\frac{5}{6}} \right](q) \omega_{MGP}(q) \right](\br)}_{v_{T_{NL}}(\br)},
\end{align}
\begin{align}
\label{energy1}
T_{NL}[\rho] =& \int \d\br \int_0^1 \d t  \rho(\br) v_{T_{NL}}[\rho_t](\br), 
\end{align}
where $\hat{F}^{-1}\left[ \right]$ indicates the inverse Fourier transform, and $v_{TF}(\br) + v_{vW}(\br)$ is the local part of the Kinetic energy potential composed of the sum of Thomas--Fermi and von Weizs\"{a}ker terms. 
To avoid spurious dependencies from the parameter $a$, we choose such boundary conditions as $\omega_{MGP}(q=0)=0$. 



From \eqn{energy1} and \eqn{eq:energy_final}, the energy expression is derived. Namely,
\begin{equation}
\label{energy}
T_{NL}[\rho] = \frac{3}{5} \int \d\br \rho(\br) v_{T_{NL}}(\br).
\end{equation}

The simplicity of the above equation results from double integration of the kernel in \eqn{eq:vt}. A similar expression would be recovered for the arithmetically symmetrized kernel, while a more complex expression involving a double integration of the inverse Lindhard function would be required for the geometrically symmetrized kernel. 


We mention here that the so-called LQ and HQ functionals \cite{chai2007} (their potentials are constructed {\it ad hoc} to exactly reproduce the high-$q$ or the low-$q$ limits) feature a nonlocal kinetic energy that is also evaluated with a line integral similar to \eqn{energy1}. The line integral in Ref.\ \citenum{chai2007}, however, is carried out {only for the nonhomogeneous component of the density} ({\it i.e.}, $\rho(\br) - \rho_0$). 

\section{Computational details}
The OF-DFT calculations are carried out  with both a modified version of ATLAS\cite{ATLAS}  and PROFESS~3.0\cite{PROFESS3}.  Calculations involving the HC KEDF are performed with PROFESS~3.0, all others are carried out in ATLAS and double checked with PROFESS~3.0 for consistency. The benchmark KS-DFT calculations are carried out with the same local pseudopotentials as the OF-DFT simulations. We employed the optimal effective local pseudopotentials (OEPP) \cite{OEPP} and the bulk-derived local pseudopotentials (BLPS) of Huang \& Carter \cite{Huang_2008}. KS-DFT calculations with OEPP are performed with Quantum-ESPRESSO (QE) \cite{qe}, while the ones employing BLPS are carried out with ABINIT \cite{abinit}. In all calculations, the LDA xc functional by Perdew and Zunger \cite{Perdew1981} is adopted.

Crystal diamond (CD), body-centered cubic (BCC) and face-centered cubic (FCC) Silicon phases, as well as nine III-V ZB semiconductors  are selected as benchmark systems. We compute total energy curves as a function of the lattice constant and extract the cell volume, $V_0$, the minimum energy, $E_0$ and the bulk modulus, $B_0$. These were computed using the prescription of Carter and coworkers \cite{huan2010} fitting the energy curves vs volume against Murnaghan's equation of state \cite{Murnaghan_1944} within 0.95 $V_0$ to 1.05 $V_{0}$.

In all OF-DFT calculation, the grid spacing of 0.2 \AA\ in ATLAS and Kinetic energy cutoffs 1600 eV in PROFESS~3.0 are sufficient for well-converged total energies (1 meV/cell). In KS-DFT simulations, for bulk properties the $20\times 20\times 20$ $k$-points meshes and 800 eV Kinetic energy cutoffs are used. To obtain smooth electron density, a denser grid (larger Kinetic energy cutoffs) are adopted for both OF-DFT and KS-DFT calculations to keep the real space electron density are represented on $54\times 54\times 54$ for ZB GaAs and $36\times 36\times 36$ for CD Silicon.

\section{Results and Discussion}
We have benchmarked MGP against KS-DFT and compared it to WT, WGC and HC KEDFs for metallic and semiconducting phases of Si, as well as the ZB phase of nine II-V semiconductors. In the following, we will only compare the so-called ``optimal'' parameters for all KEDFs, including MGP. We should remark that the optimal parameters are pseudopotential dependent. Thus, we optimize all KEDFs parameters accordingly. With exception of WT, all other KEDFs have two adjustable parameters:
 MGP has $a$ and $b$, defining the Kinetic electron term; 
 WGC has the effective electron density, $\rho^*$, and the averaged Fermi wavevector parameter, $\gamma$; and 
 HC has $\beta$ and $\lambda$, defining the long-range portion of the functional.


\subsection{Bulk properties for both metallic and semiconductor phases of Silicon}
\begin{table}[!h]
\caption{\label{tab:Si} Equilibrium volume ($V_0$ in Bohr$^3$), minimum energy ($E_0$ in eV/atom), and bulk modulus ($B_0$ in GPa) for metallic phases (BCC,FCC) and semiconductor phase (CD) of silicon computed by OF-DFT/BLPS and OF-DFT/OEPP using the MGP, WGC, WT, and HC KEDFs with optimal choices of their parameters. The corresponding benchmark KS-DFT/BLPS and KS-DFT/OEPP results are also given. HC KEDF values are taken from Ref.\citenum{huan2010}. WGC's optimal parameters are taken from Ref.\citenum{shin2014} and \citenum{zhou2005}.}
{\fontsize{11}{11}\selectfont
\begin{tabular}{lcccccc}
\hline \hline
Systems & Functionals &Pseudopotentials& Parameters    & $V_0(a.u/Cell)$    &$E_0(eV)$    &    $B_0(GPa)$    \\
\hline\\[-12pt]
FCC&KS    & OEPP &-- --         &118.2    & -107.761  & 71\\
  &MGP    & OEPP &a=0.5,b=0.3   &119.5    & -107.765    & 73 \\
  &WGC    & OEPP &$\gamma=2.2,\rho*=\rho_{0}$ &112.7    & -107.733    &113 \\
  &WGC    & OEPP &$\gamma=1.0,\rho*=1.05\rho_{0}$ & 115.9 & -107.790 & 98 \\
\hline
  &KS     & BLPS &-- --         &97.0   & -109.248  & 83 \\
  &MGP    & BLPS &a=0.6,b=0.4   &97.6   & -109.243  & 75 \\
  &WT     & BLPS &-- --         &97.5   & -109.260  & 58 \\
  &WGC    & BLPS &$\gamma=2.2,\rho*=\rho_{0}$ &94.4   & -109.258  & 96 \\
\hline\\[-12pt]
BCC&KS    & OEPP &-- --          &234.7    & -215.490  & 74 \\
  &MGP    & OEPP &a=0.66,b=0.38  &235.7    & -215.492    & 89 \\
  &WGC    & OEPP &$\gamma=2.2,\rho*=1.05\rho_{0}$ &228.9    & -215.504    & 106 \\
  &WGC    & OEPP &$\gamma=1.0,\rho*=1.05\rho_{0}$ &230.5    & -215.521  & 101 \\
\hline
  &KS     & BLPS &-- --          &197.1    & -218.556  & 98 \\
  &MGP    & BLPS &a=0.94,b=0.6    &205.4    & -218.553    &97 \\
  &WT     & BLPS &-- --          &204.3    & -218.666    & 66 \\
  &WGC    & BLPS &$\gamma=2.2,\rho*=1.05\rho_{0} $  &196.9    & -218.553    &106 \\
  \hline
CD&KS     & OEPP &-- --          &309.3  & -215.538  & 61 \\
  &MGP    & OEPP &a=0.341,b=0.45 &310.0  & -215.534  & 54 \\
  &WGC    & OEPP &$\gamma=4.2,\rho*=1.05\rho_{0}$ &352.2& -215.512 & 34\\
  &WGC    & OEPP &$\gamma=5.0, \rho*=1.05\rho_{0}$ &332.7& -214.974 & 44 \\
\hline
  &KS     & BLPS &-- --          &266.9  & -219.258  & 98 \\
  &MGP    & BLPS &a=0.364,b=0.57 &265.6  & -219.258     & 95 \\
  &HC     & BLPS &$\lambda=0.01,\beta=0.65$ &269.4     & -219.248  & 97 \\
  &WT     & BLPS &-- --          &--     &--         &-- \\
  &WGC    & BLPS &$\gamma=4.2,\rho*=1.05\rho_{0}$ &287.4 &-218.838 & 64\\
  \hline
\end{tabular}}
\end{table}


The optimal parameters for MGP functionals along with the optimal $\gamma$ and $\rho^*$ for WGC, optimal $\beta$ and $\lambda$ for HC functionals and the calculated quantities for all KEDFs are collected in Table \ref{tab:Si}.

Since BCC and FCC structures of Si are metallic, we expect that WGC KEDF produces more or comparablely accurate bulk properties compared to HC KEDFs. Thus, for these metallic phases we just compare MGP results to WGC, WT, and the benchmark KS-DFT results.  Comparing to KS-DFT, MGP reproduces bulk properties overall improving on WT and WGC KEDFs with both OEPP and BLPS. This is especially the case for the total MGP energies which deviate from KS-DFT by less than 5 meV/cell in all cases.

 Modelling CD Si with OF-DFT has historically been a challenge which has been addressed by several nonlocal functionals with density-dependent kernel \cite{wang1998,wang1999,shin2014,huan2010}. For example \cite{zhou2005}, WT is unable to reproduce a bound curve for CD Si. As MGP functional has a density-independent kernel, we had no expectations that CD Si would be modeled correctly. Table \ref{tab:Si}, however, shows that MGP is capable of producing bound energy curves and overall bulk properties that are close to the KS-DFT benchmark compared to other KEDFs. To the best of our knowledge, MGP is the only KEDF with density-independent kernel capable of simultaneously reproducing KS-DFT equilibrium total energies, bulk moduli, and equilibrium volumes for both metallic and semiconductor phases of Silicon. 

We also tested a modified version of MGP only including WT and Kinetic electron term. We found that the modified functional could not reproduce the benchmark despite efforts in optimizing the $a$ and $b$ parameters. This indicates that, to achieve accurate results it is not enough to simply add the correction terms separately. Instead, the correction term derived from functional integration and the Kinetic electron must be included together for MGP to approach both metallic and semiconducting phases.  
\subsection{Benchmarks for III-V semiconductors}
In this section, we present benchmark MGP against KS-DFT for binary III-V semiconductors in the ZB phase. These are challenging systems for KEDFs. For example, it was shown \cite{huan2010,zhou2005,xia2012} that WGC is not appropriate and that proper long-range behavior of the kernel needs to be included. We have developed MGP with semiconductors and finite systems in mind and, specifically, the Kinetic electron term aims at correcting the KEDF kernel in its long-range (low $q$) behavior.
\begin{table}[!h]
\caption{\label{tab:Semiconductors} Equilibrium volume ($V_0$ in Bohr$^3$), minimum energy ($E_0$ in eV/atom), and bulk modulus ($B_0$ in GPa) for various ZB III-V semiconductors computed with OF-DFT/BLPS and the MGP and HC functionals with optimal choices of their parameters. The benchmark KS-DFT/BLPS results are also given.   KS-DFT/BLPS and OF-DFT/BLPS with HC functional values are taken from Ref.\citenum{huan2010}.}
{\fontsize{11}{11}\selectfont
\begin{tabular}{lccccc}
\hline \hline
Systems & Functional & Parameters    & $V_0(a.u/Cell)$    &    $E_0(eV)$    &    $B_0(GPa)$    \\
\hline
GaAs    &KS    &                               & 274.2  & -235.799    & 75 \\
        &HC    & $\lambda=0.0130,\beta=0.783$ & 275.3    & -235.782    & 81 \\
        &MGP& $a=0.434,b=0.524$            & 275.2  & -235.801    & 75 \\
\hline
GaSb    &KS &                               & 354.2    & -209.697    & 56 \\
        &HC    & $\lambda=0.0100,\beta=0.720$ & 355.5    & -209.739    & 58 \\
        &MGP& $a=0.341,b=0.752$               & 360.5  & -209.696    & 49 \\
\hline
GaP        &KS &                               & 254.0    & -243.079    & 80 \\
        &HC & $\lambda=0.0100,\beta=0.791$ & 255.0    & -243.057    & 87 \\
        &MGP& $a=0.405,b=0.394$               & 252.5  & -243.077    & 82 \\
\hline
AlAs    &KS &                              & 294.3  & -232.908    & 80 \\
        &HC & $\lambda=0.0125,\beta=0.825$ & 301.6    & -232.912    & 76 \\
        &MGP& $a=0.396,b=0.424$            & 296.2    & -232.903    & 76 \\
\hline
AlP        &KS &                               & 274.2    & -240.182    & 90 \\
        &HC &$\lambda=0.0120,\beta=0.845$  & 272.8    & -240.199    & 91 \\
        &MGP& $a=0.378,b=0.319$            & 273.0    & -240.180    & 81 \\
\hline
AlSb    &KS &                               & 382.0    & -206.606    & 60 \\
        &HC &$\lambda=0.0120,\beta=0.750$  & 377.9    & -206.588    & 61 \\
        &MGP& $a=0.359,b=0.832$            & 383.9    & -206.606    & 55 \\
\hline
InP        &KS &                               & 310.7    & -235.722    & 73 \\
        &HC &$\lambda=0.0120,\beta=0.885$  & 309.4    & -235.696    & 66 \\
        &MGP& $a=0.394,b=0.351$               & 308.2    & -235.724    & 62 \\
\hline
InAs    &KS &                               & 331.5  & -228.537    & 65 \\
        &HC &$\lambda=0.0142,\beta=0.875$  & 334.0    & -228.523    & 63 \\
        &MGP& $a=0.408,b=0.444$            & 330.1  & -228.532    & 61 \\
\hline
InSb    &KS &                               & 424.5    & -202.387    & 50 \\
        &HC &$\lambda=0.0120,\beta=0.810$  & 424.1    & -202.381    & 49 \\
        &MGP& $a=0.361,b=0.850$               & 426.9    & -202.386    & 46 \\
\hline \hline
\end{tabular}}
\end{table}

Table \ref{tab:Semiconductors} lists the performance of MGP compared to HC as well as KS-DFT. The results indicate that both MGP and HC functional can reproduce accurate equilibrium volume and bulk modules. The predicted equilibrium volumes with MGP and HC lie within 2\% of the KS-DFT results for all considered semiconductors. For the bulk moduli, MGP and HC are within 10 GPa compared to KS-DFT. MGP total equilibrium energies are within 5 meV/cell to the KS-DFT benchmark. This improves upon HC which finds itself within 42 meV/cell. Overall, however, the quality of the calculated MGP energies is similar to HC for the considered systems. 

\subsection{Analysis of the electron density}
\begin{figure}[htp]
\begin{center}
\includegraphics[width=0.8\textwidth]{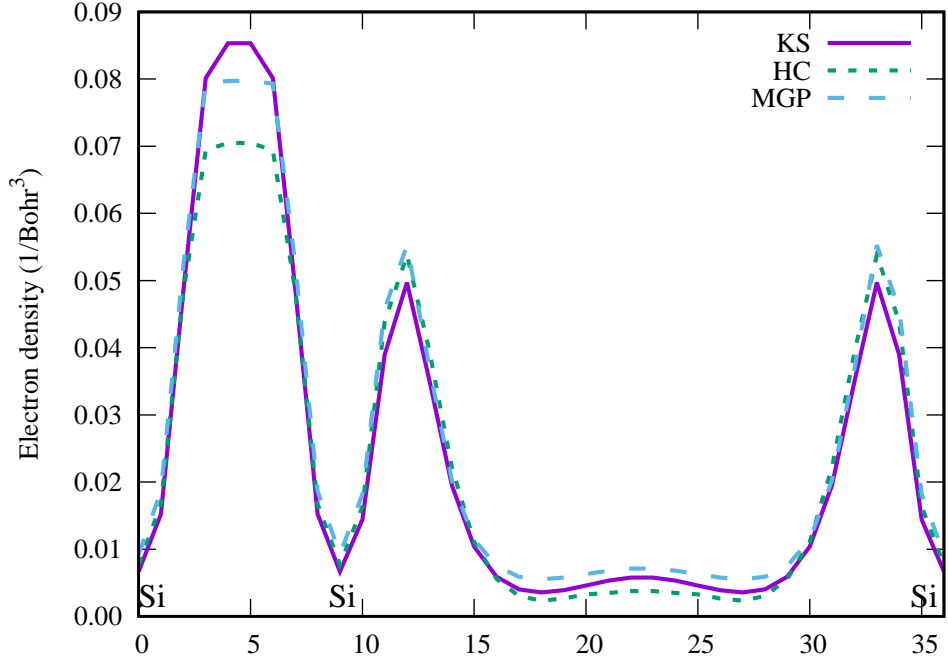}
\end{center}
\caption{\label{Si_Density} The electron density of CD Silicon along the [111]  direction. Purple line: KS-DFT. Green line: OF-DFT with HC functional. Blue line: OF-DFT with MGP functional. Vertical axis is electron density in 1/ Bohr$^{3}$; horizontal axis represents the grid points and the position of the Silicon atoms is indicated with a ``Si'' label in the graph.}
\end{figure}

\begin{figure}[htp]
\begin{center}
\includegraphics[width=0.8\textwidth]{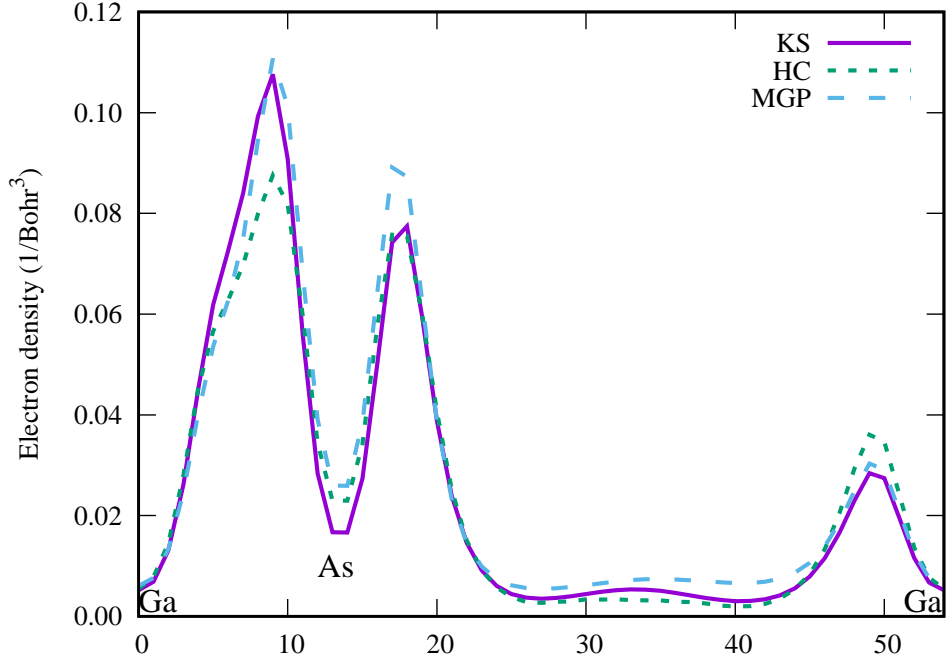}
\end{center}
\caption{\label{GaAs_Density} The electron density of ZB GaAs along the [111]  direction. Purple line: KS-DFT. Green line: OF-DFT with HC functional. Blue line: OF-DFT with MGP functional. Electron densities are in atomic units; horizontal axis represents the grid points and the position of the Silicon atoms is indicated with a ``Si'' label in the graph.}
\end{figure}

We have also evaluated  MGP by comparing the ground state electron densities produced by OF-DFT with MGP and HC KEDFs, as well as KS-DFT for CD Si and ZB GaAs. Once again, we employ optimal parameters  for the energy functionals. Figure \ref{Si_Density} and \ref{GaAs_Density} display the electron density distribution along the [111] direction for CD Silicon and ZB GaAs, respectively. Inspection of the figures reveals that the MGP electron density reproduces KS-DFT quite well. While in both cases HC tends to underestimate the density in the bonding region, MGP is much closer to KS-DFT. Conversely, in the space between non-covalently bonded atoms, HC slightly underestimates the KS-DFT density while MGP slightly overestimates it. 

We have tested several functionals for their performance in reproducing KS-DFT electron densities for these systems and found that WT does a remarkably good job. In this respect, because MGP is based on WT, it inherits some of its good traits. 


\subsection{Computational efficiency}
\begin{table}[!h]
\caption{\label{tab:time} Total computational wall time, number of truncated Newton (TN) optimization steps, total number of calls for evaluation of potential, wall time per call in evaluation of the total potential, and wall time per call in evaluation of kinetic potential with the WGC, MGP, and WT KEDFs. The wall time for evaluation of MGP kernel are also listed.}
\begin{tabular}{lccccc}
\hline \hline \\ [-12pt]
                                          &  KEDFs &2 atoms   &16 atoms  &128 atoms & 1024 atoms \\
\hline
Total Wall time [s]                       &  WGC   & 8.7      & 87.1     &830.4   & 6264.2  \\
                                          &  MGP   & 8.9      & 76.4     &648.4   & 5486.4 \\
                                          &  WT    & 5.0      & 51.5     &456.8   & 4342.3 \\
\hline
Number of optimized steps                 &  WGC   & 10       &10        &11      & 12\\
                                          &  MGP   & 8        & 8        &10      & 11 \\
                                          &  WT    & 7        & 8        &9       & 11 \\
\hline
Number of calls for evaluation potentials & WGC    &248       &253       &280     & 282  \\
                                          & MGP    &191       &212       &224     & 228  \\
                                          & WT     &173       &207       &219     & 236  \\
\hline
Wall time of total potential[s/call]      & WGC    &0.03      &0.32      &2.81    & 21.27    \\
                                           & MGP    &0.01      &0.16      &1.42    & 12.64    \\
                                          & WT     &0.02      &0.23      &1.97    & 17.3    \\
\hline
Wall time of kinetic potential[s/call]    & WGC    &0.03      &0.27      &2.37    & 17.88    \\
                                           & MGP    &0.01      &0.12      &1.02    & 9.00    \\
                                          & WT     &0.02      &0.19      &1.55    & 13.71    \\
\hline
Wall time of evaluation kernel[s]         & MGP    &4.9       &37.8      &299.9   & 2383.3  \\
\hline
\end{tabular}
\end{table}
To evaluate the computational efficiency of MGP, we compare the computational cost for optimizing the electron density in CD Si (2 atoms), $2\times 2\times 2$ (16 atoms), $4\times 4\times 4$ (128 atoms), and $8\times 8\times 8$ (1024 atoms) supercells with WGC, WT, and MGP KEDFs. All calculations are performed on single thread with PROFESS employing the BLPS pseudopotentials using a large Kinetic energy cutoff (4000 eV). For all KEDFs we employ optimal parameters (WGC: $\gamma=4.2$, $\rho^{*}$=$\rho_{0}$; WT: $\alpha=\beta=5/6$; MGP: $a$=0.364 and $b$=0.570 ). Since the computational cost of HC KEDF is much higher than that of other functionals (about 1000 times higher than MGP for CD Si) \cite{huan2010,shin2014}, we exclude it in the comparison. In principle, the only difference in computational cost between MGP and WT KEDFs is the initial kernel building step in reciprocal space. 

As showed in Table \ref{tab:time}, the wall time of computing MGP kernel is about 200 times larger than a single call for the evaluation of the potential. Luckily, the kernel is evaluated only once at the beginning of the simulation. We recall that the computational cost in the initial MGP kernel is strictly linear scaling with system size instead of quasilinear (i.e., $N\ln(N)$ scaling with $N$ being the number of real-space grid points) in the evaluation of the potential. As a result, for small systems the total wall time for optimizing electron density with MGP KEDF is always between WT and WGC's wall times. Increasing the number of grid points (or the size of the system), the wall time should approach WT  (although this should be system dependent). 

Among WGC, MGP and WT KEDFs, WGC is the most expensive KEDF in terms of total wall time. Additionally, we find that MGP KEDFs converges in all cases considered with 12 iterations of a truncated Newton minimization. In contrast, WGC KEDF is numerical unstable in some cases (this is a well-known limitation of this functional \cite{huan2010,shin2014}).

In summary, the computational efficiency of MGP closely resembles WT's and both MGP and WT are more computationally efficient than WGC. 

\section{Conclusions}

We have formulated and implemented MGP, a new nonlocal Kinetic energy functional with a {\it density independent kernel}. In MGP, the inverse response function of the FEG is functional-integrated to yield a new kernel. In a second step, the kernel is augmented by a ``Kinetic electron'' which is opposite to the exchange hole. 

Our pilot calculations show that MGP improves dramatically over currently available nonlocal functionals with density independent kernels and performs even better than the best available functionals  (featuring density dependent kernels) for both metallic and semiconducting phases of Si as well as the ZB phases of nine common III-V semiconductors. 
Although the results presented here are quite encouraging, we should take them with a grain of salt. In this work we have only considered bulk systems, and the lingering question is: can MGP deliver similar quality results for finite systems or systems with vacancies? Initial tests of MGP applied to finite systems (isolated clusters) are quite encouraging and will be the subject of a follow-up work.

In light of recent debates criticizing the use of the total electronic energy as the only descriptor relevant for optimizing a density functional \cite{Medvedev_2017}, we have also inspected the ability of MGP to reproduce electron density distributions. We find that MGP performs quite remarkably in this respect. For example, MGP's densities are much closer to KS-DFT in bonding regions for CD Silicon and ZB GaAs compared to the current state-of-the-art HC functional. Interestingly, we find that although for semiconductors and nonsimple metals WT is unable to reproduce observables related to the energy, it delivers quite accurate electron densities. As MGP derives from WT, this is probably the reason why MGP delivers accurate densities.
  
Finally, the benchmarks for computational cost indicate that our new KEDF is almost as computationally efficient as WT which is of similar scaling as GGA functionals.  

Our analysis shows that there is room for improvement, particularly for two aspects of the MGP functional. First, the Kinetic electron can be better parametrized following prescriptions that have been very successful in formulating exchange functionals in real space \cite{perd1985,Gritsenko_1995}. Secondly, there exist formulations of ``jellium with gap'' models for the Lindhard function \cite{constantin2017} which have been recently (and successfully) applied to the formulation of GGA KEDF \cite{Smiga_2017}. These can be extended to the line-integral formulation of MGP. Thirdly, a density-dependent kernel version of MGP can be constructed to satisfy additional exact conditions, such as density and coordinate scaling relations \cite{levy1988}.

\section{Supplementary Material}
The supplementary material section includes an analysis of different ways of symmetrizing the kernel with respect to the spatial coordinates and the respective electron density dependence.

\section{Acknowledgments}
This material is based upon work supported by the National Science Foundation under Grant No. CHE-1553993.
We thank Emily Carter and her group members for insightful discussions.

\end{document}